\documentclass{IOS-Book-Article}
\usepackage{mathptmx}
\usepackage{xspace}
\usepackage{url}
\usepackage{color}
\usepackage{multirow}
\usepackage{graphicx}
\usepackage{mathtools}

\newcommand{\CPTWOK}{{\tt CP2K}\xspace}
\newcommand{\DBCSR}{{\tt DBCSR}\xspace}
\newcommand{\LIBXSMM}{{\tt LIBXSMM}\xspace}
\newcommand{\LIBCUSMM}{{\tt LIBCUSMM}\xspace}

\def\hb{\hbox to 10.7 cm{}}

\begin{document}

\pagestyle{headings}
\def\thepage{}

\begin{frontmatter}              % The preamble begins here.

\title{\DBCSR: A Blocked Sparse Tensor Algebra Library}

\markboth{}{September 2019\hb}

\author[A]{\fnms{Ilia} \snm{Sivkov}\thanks{E-mail: ilia.sivkov@chem.uzh.ch}},
\author[A]{\fnms{Patrick} \snm{Seewald}\thanks{E-mail: patrick.seewald@chem.uzh.ch}},
\author[A]{\fnms{Alfio} \snm{Lazzaro}\thanks{E-mail: alazzaro@cray.com, now at Cray Switzerland GmbH, Switzerland}},
and
\author[A]{\fnms{J{\"u}rg} \snm{Hutter}\thanks{E-mail: hutter@chem.uzh.ch}}

\address[A]{University of Zurich, Department of Chemistry, Switzerland}

\begin{abstract}
Advanced algorithms for large-scale electronic structure calculations are mostly based on processing multi-dimensional sparse data. Examples are sparse matrix-matrix multiplications in linear-scaling Kohn-Sham calculations or the efficient determination of the exact exchange energy. When going beyond mean field approaches, e.g. for Moller-Plesset perturbation theory, RPA and Coupled-Cluster methods, or the GW methods, it becomes necessary to manipulate higher-order sparse tensors. Very similar problems are also encountered in other domains, like signal processing, data mining, computer vision, and machine learning. 
With the idea that the most of the tensor operations can be mapped to matrices, we have implemented sparse tensor algebra functionalities in the frames of the sparse matrix linear algebra library \DBCSR (Distributed Block Compressed Sparse Row). 
\DBCSR has been specifically designed to efficiently perform blocked-sparse matrix operations, so it becomes natural to extend its functionality to include tensor operations. 
We describe the newly developed tensor interface and algorithms. In particular, we introduce the tensor contraction based on a fast rectangular sparse matrix multiplication algorithm. 
\end{abstract}

\end{frontmatter}
\markboth{September 2019\hb}{September 2019\hb}

\section{Introduction}
\label{sec:Introduction}

Most, if not all the modern scientific simulation packages utilize matrix algebra operations. Often, due to the nature of simulated systems, the structure of matrices and tensors is sparse with a low degree of nonzero elements ($<10\%$). 
Applications exploiting the sparsity include the linear scaling density functional theory~\cite{joost1M}, cubic scaling RPA algorithm and a similar approach to fast, quadratic scaling Hartree-Fock exchange~\cite{cubic-rpa} in the quantum chemistry \CPTWOK framework~\cite{cp2k}. 
The first method works with sparse matrices, while the other two algorithms rely on contractions involving sparse 3-rank tensors. 
Due to the nature of the studied chemical systems, this naturally leads to a blocked sparsity pattern, with chemically motivated block sizes.
Therefore, the implementation of such methods requires convenient and effective tools and libraries to work also with block-sparse matrices and tensors, with a range of occupancy between $0.01\%$ up to dense.

The highly optimized linear algebra library \DBCSR (Distributed Block Compressed Sparse Row) has been specifically designed to efficiently perform block-sparse and dense matrix operations, covering a range of occupancy between $0.01\%$ up to dense. It is parallelized using MPI and OpenMP, and can exploit GPU accelerators using CUDA. The more detailed description of these features can be found in the previous works~\cite{dbcsr, ole, Lazzaro:2017:IES:3093172.3093228}. Here we give an overview of the library in section~\ref{sec:DBCSR}.

Although \DBCSR supports multiplications of rectangular matrices, the implemented algorithm was inefficient whenever the resulting matrix has a much smaller size than input matrix sizes ($<1000$ smaller). This matrix multiplication can be used for the realization of tensor contraction since the tensor contraction can be mapped to matrix-matrix multiplications \cite{solomonik2015communication}. In section~\ref{sec:rectangular} we present an optimized implementation for such a case.
Additionally, we have developed the tensor algebra operations as an extension of the \DBCSR library. In section~\ref{sec:tensor} we present an overview of the new functionalities.
The main operation which is used in tensor algebra is a contraction between two tensors over a set of indices. In many of methods, the rank of tensors is no more than 4 and therefore the non-trivial contractions can be performed over 1-3 indices. Finally, section~\ref{sec:conclusion} reports the conclusion.

\subsection{Related Work}

Other implementations of tensor libraries are described in Ref.~\cite{lewis2015clustered, solomonik2015sparse, JCC:JCC23377, rajbhandari2013framework, landry}, while Ref.~\cite{Kolda:2009:TDA:1655228.1655230} presents an overview of tensor algebra applications. The proposed tensor library implementation in \DBCSR differs from these implementations since it is specifically targeting block-sparse tensor contractions with a wide range of occupancy between $0.01-10\%$ by optimally exploiting block sparsity. Existing parallel sparse tensor libraries have limited parallel scalability \cite{JCC:JCC23377}, do not prove to be more efficient compared to the dense case \cite{lewis2015clustered}, or have low sequential efficiency \cite{solomonik2015sparse}. For matrix-matrix multiplications, the \DBCSR library already provides an efficient and scalable solution without the above-mentioned shortcomings.

\section{{The \large \textbf{\DBCSR}} Library}
\label{sec:DBCSR}

\DBCSR is written in Fortran and is freely available under the GPL license from \url{https://github.com/cp2k/dbcsr}.
\DBCSR matrices are stored in a blocked compressed sparse row (CSR) format distributed over a two-dimensional grid of $P$ MPI processes. Matrix-matrix multiplication is implemented by means of the Cannon algorithm~\cite{cannon}. As part of this work, two novel implementations are specifically introduced for rectangular matrix multiplications similar to one iteration of CARMA algorithm~\cite{carma} (see section~\ref{sec:rectangular}) and for the tensor contraction algorithm (see section~\ref{sec:tensor}). The latter uses the same idea as for the rectangular matrix multiplication with a slightly different implementation. 

In the Cannon algorithm, only the matrices $A$ and $B$ are communicated for the multiplication $C = C + A\times B$. The amount of communicated data by each process scales as $\mathcal{O} (1/\sqrt{P})$. These communications are implemented with asynchronous point-to-point MPI calls, using the MPI Funneled mode~\cite{Lazzaro:2017:IES:3093172.3093228}. The local multiplication will start as soon as all the data has arrived at the destination process, and it is possible to overlap the local computation with the communication if the network allows that. 

The local computation consists of pairwise multiplications of small dense matrix blocks, with dimensions $(m \times k)$ for $A$ blocks and $(k \times n)$ for $B$ blocks. It employs a cache-oblivious matrix traversal to fix the order in which matrix blocks need to be computed, in order to improve memory locality. First, the algorithm loops over $A$ matrix row-blocks and then, for each row-block, over $B$ matrix column-blocks. 
Then, the corresponding multiplications are organized in batches. 
Multiple batches can be computed in parallel on the CPU by means of OpenMP threads or alternatively executed on a GPU.  A static assignment of batches with a given $A$ matrix row-block to threads is employed in order to avoid race conditions.  
Processing the batches has to be highly efficient. For this reason, specific libraries were developed that outperform vendor BLAS libraries, namely \LIBCUSMM for GPU and \LIBXSMM for CPU/KNL systems~\cite{libxsmm, parco_knl}.

For GPU execution, data is organized in such a way that the transfers between the host and the GPU are minimized. A double-buffering technique, based on CUDA streams and events, is used to maximize the occupancy of the GPU and to hide the data transfer latency~\cite{ole}.
When the GPU is fully loaded, the computation may be simultaneously done on the CPU. 
\LIBCUSMM employs an auto-tuning framework in combination with a machine learning model to find optimal parameters and implementations for each given set of block dimensions. 
For a multiplication of given dimensions $(m, n, k)$, \LIBCUSMM's CUDA kernels are parametrized over $7$ parameters, affecting:
\begin{itemize}
    \item algorithm (different matrix read/write strategies)
    \item amount of work and number of threads per CUDA block
    \item number of matrix element computed by one CUDA thread
    \item tiling sizes
\end{itemize}
yielding $\approx 30{,}000$~-~$150{,}000$ possible parameter combinations for each of about $\approx 75{,}000$ requestable $(m, n, k)$-kernels. These parameter combinations result in vastly different performances. We use machine learning to derive a performance model from a subset of tuning data that accurately predicts performance over the complete kernel set. The model uses regression trees and hand-engineered features derived from the matrix dimensions, kernel parameters, and GPU characteristics and constraints. 
To perform the multiplication the library uses Just-In-Time (JIT) generated kernels or dispatches the already generated code.
In this way, the library can achieve a speedup in the range of 2--4x with respect to batched DGEMM in cuBLAS. 

\DBCSR operations include sum, dot product, and multiplication of matrices, and the most important operations on single matrices, such as transpose and trace. Additionally, the library includes some of the linear algebra methods, such as the sign matrix algorithm~\cite{joost1M} and matrix inverse. These methods were ported from \CPTWOK to \DBCSR. The sign matrix algorithm is used in the linear scaling density functional theory in order to find a ground state of the quantum systems. As associated methods, we have ported the matrix-vector multiplication operation and an interface to some SCALAPACK operations.

\section{Optimized Rectangular Matrix Multiplication Algorithm and Implementation}
\label{sec:rectangular}

\begin{figure}
\centering
\includegraphics[scale=0.7]{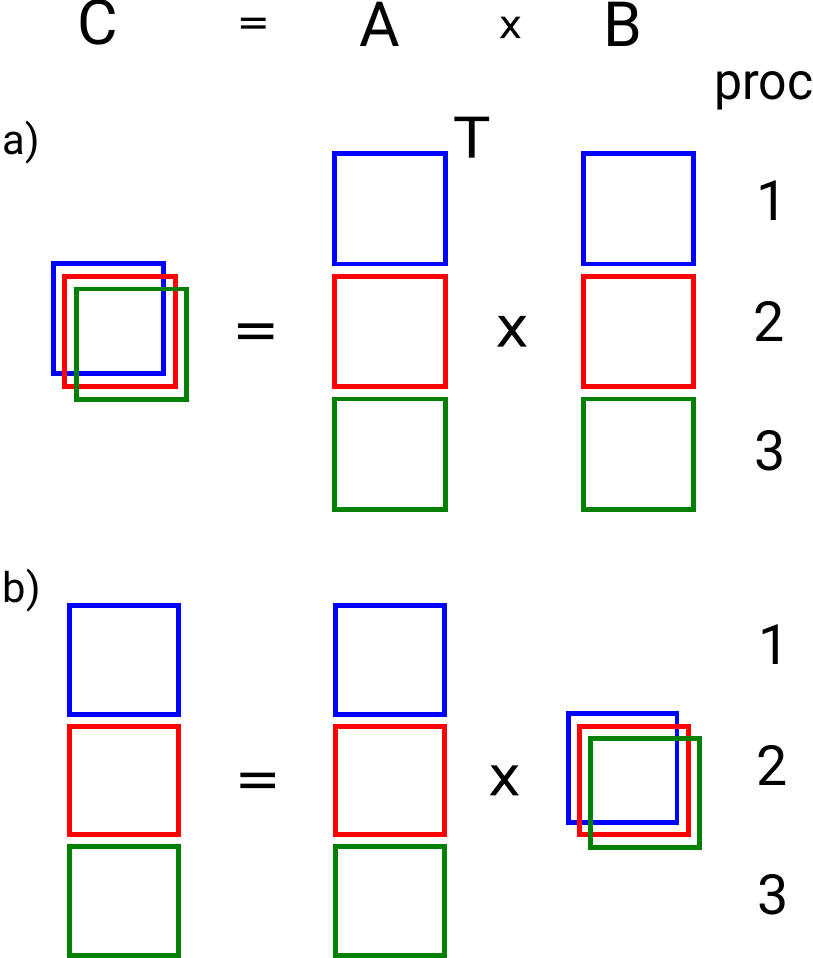}
\caption{Communication-avoiding algorithm for the rectangular matrix-matrix multiplication.
a) Middle dimension $K$ is the largest (case 1), $C$ is replicated and $A$ and $B$ are distributed on a linear grid. b) Outer dimension $M$ is the largest (case 2), $C$ and $A$ are distributed on a linear grid and $B$ is cloned or distributed.}
\label{fig:carma}
\end{figure}

Despite the Cannon algorithm gives in general good performance for the sparse matrix multiplication of any size, it loses its efficiency in the case where the size of the resulting matrix $C$ ($S_C = O_C M N$) is much smaller than the sizes of the input $A$ ($S_A = O_A M K$) and/or $B$ ($S_B = O_B K N$) matrices, where $M,N,K$ are the dimensions of the dense matrices and $O_A, O_B, O_C$ their occupancy values. This is a direct consequence of the algorithm since it requires the communication of $A$ and $B$ data on a 2D grid of $P$ processors, while $C$ remains local to each processor. In particular, for the multiplication of two rectangular matrices the Cannon algorithm requires to communicate per each processor~\footnote{Here we assume a uniform distribution of the non-zero elements in the matrices without losing generality.}
\begin{equation}
    T_{w} = \frac{S_A+S_B}{\sqrt{P}} = \frac{K(O_AM+O_BN)}{\sqrt{P}}.
    \label{eq:cannon}
\end{equation} 
Therefore, the communication will be dominated by one of the dimension whenever is much larger than the other two. We can distinguish the two important cases:
\begin{enumerate}
    \item $M \ll K$ and $N \ll K$, which corresponds to $S_C \ll \{S_A, S_B\}$
    \item $K \ll M$ and $K\le N$, which corresponds to $S_B \ll \{S_A, S_C\}$
\end{enumerate}
According to Ref.~\cite{demmel2013communication}, a communication-optimal algorithm for this case is obtained by dividing the original matrix multiplication into smaller tasks such that each task is local to a process subgroup. Inspired by this idea, we redistribute the matrices on a linear MPI grid (see Figure~\ref{fig:carma}) and perform the multiplication locally. 
We describe the implementations for the two cases in the following subsections. We also report the results of some tests we performed for a variety of matrix, block sizes and occupancy values of our interests ($10\%-50\%$ often present in \CPTWOK). We used double precision matrices with sizes of the order $M,N=\mathcal{N},\ K=\mathcal{N}^2$ and $\mathcal{N} = 10^3$. The calculations were performed using the Cray XC50 ``Piz Daint'' supercomputer at the Swiss National Supercomputing Centre (CSCS). Each node of the system is equipped by a CPU Intel Xeon E5-2690 v3 @ 2.60GHz (12 cores, 64GB DRAM) and a GPU NVIDIA Tesla P100 (16GB HBM). For the MPI configuration, we used 1 rank per node and 12 OpenMP threads per rank. Each multiplication was performed 100 times to exclude the fluctuations of performance due to hardware glitches.

\subsection{$S_C \ll \{S_A, S_B\}$ } 
\begin{figure}
\centering
\includegraphics[scale=0.7]{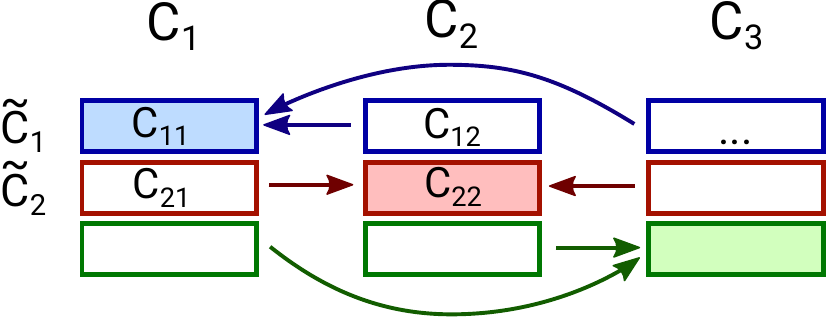}
\caption{Reduce operation of the matrix $C$ after the local multiplication when $S_C \ll \{S_A, S_B\}$.}
\label{fig:allreduce}
\end{figure}

Matrices $A$ and $B$ are redistributed on a linear MPI grid and the $A$ matrix is transposed, such as the longest dimension $K$ is now distributed over the $P$ processors (see Figure~\ref{fig:carma}a). Then a local multiplication is executed, which gives $\widetilde C_i = A_i^T \cdot B_i$, with $i=1,...,P$. Here $\widetilde C_i$ corresponds to a partial result of the full, undistributed, matrix $C$. Therefore we have to reduce all  $\widetilde C_i$ and redistribute the result according to the original 2D grid distribution and sum to the input $C$ matrix to get the final distributed $C$ matrix result over the 2D grid (see Figure~\ref{fig:allreduce}). This algorithm runs in $P$ steps, where for each step we send and receive the proper $C$ data and run the local reduction. It is implemented with MPI asynchronous communications, such as we do overlap the communication of the data with the local reduction. In the end, each processor requires $S_C$ data. Including the initial redistribution of the $A$ and $B$ matrices, we get that the total amount of data communicated by each processor is:
\begin{equation}
    T_{w}^\prime = \overbracket{\left(\frac{S_A+S_B}{P}\right)}^{\text{2D } \rightarrow \text{1D grid}} + S_C.
    \label{eq:rect1}
\end{equation} 
We can now consider the ratio with the Cannon algorithm (Eq.~\ref{eq:cannon}), which leads to a reduction in communicated data $\sqrt{P}/(1+RP)$, where $R=S_C/(S_A+S_B)$. Therefore, the ratio scales as $\mathcal{O} (1/\sqrt{P})$. Finally, it is important to note that by multiplication of the sparse matrices even with high sparsity the result might be dense (so called Birthday Paradox \cite{Cormen:2001:IA:580470}). We can evaluate an upper limit on the $O_C$ by combining the Eq.~\ref{eq:cannon} and Eq.~\ref{eq:rect1} such that $T_w < T_w^\prime$:
\begin{equation}
    O_C < \frac{1}{MN} \left(T_w - \frac{S_A+S_B}{P}\right)
    \label{eq:limit1}
\end{equation} 
If we omit redistribution costs and assume that $O_A=O_B=O$ then we can write:
\begin{equation}
    \frac{O_C}{O} < \frac{K(M+N)}{MN\sqrt{P}}.
\end{equation}
As an example, for $M,N=\mathcal{N},\ K=\mathcal{N}^2$ and $\mathcal{N}=10^3,\ P=100$ we get $O_C/O < 2 \cdot 10^2$.

The results of the tests are presented in Figure~\ref{fig:ratio}a. Overall, the new implementation gives a speed-up up to 3x with respect to the Cannon algorithm for high occupancy ($>10\%$), which becomes negligible when we reach the upper limit reported in the Eq.~\ref{eq:limit1}. As expected, the speed-up decreases with the number of processors.

\subsection{$S_B \ll \{S_A, S_C\}$} 
Matrices $A$ and $B$ are redistributed in a linear MPI grid, such as the longest dimension $K$ is now distributed over the $P$ processors (see Figure~\ref{fig:carma}b). A virtual column-grid is created for the $A$ matrix to be compatible with the row-grid of the matrix $B$. Then the standard Cannon algorithm is executed over this virtual topology made of $P$ steps. Virtual column-grid does not require communication of the $A$ data and therefore only the communication of the matrix $B$ is required. Finally, $C$ result is redistributed to the original 2D grid and accumulated to the input $C$ matrix. The total amount of communicated data by each processor is:
\begin{equation}
    T_{w}'' = \overbracket{\left(\frac{S_A+S_B+S_C}{P}\right)}^{\text{2D } \rightarrow \text{1D} \rightarrow \text{2D grid}} + S_B.
    \label{eq:rect2}
\end{equation} 
Also in this case the ratio of the communicated data with respect to Cannon implementation scales as $\mathcal{O} (1/\sqrt{P})$.

\begin{figure}
\centering
\includegraphics[scale=0.7]{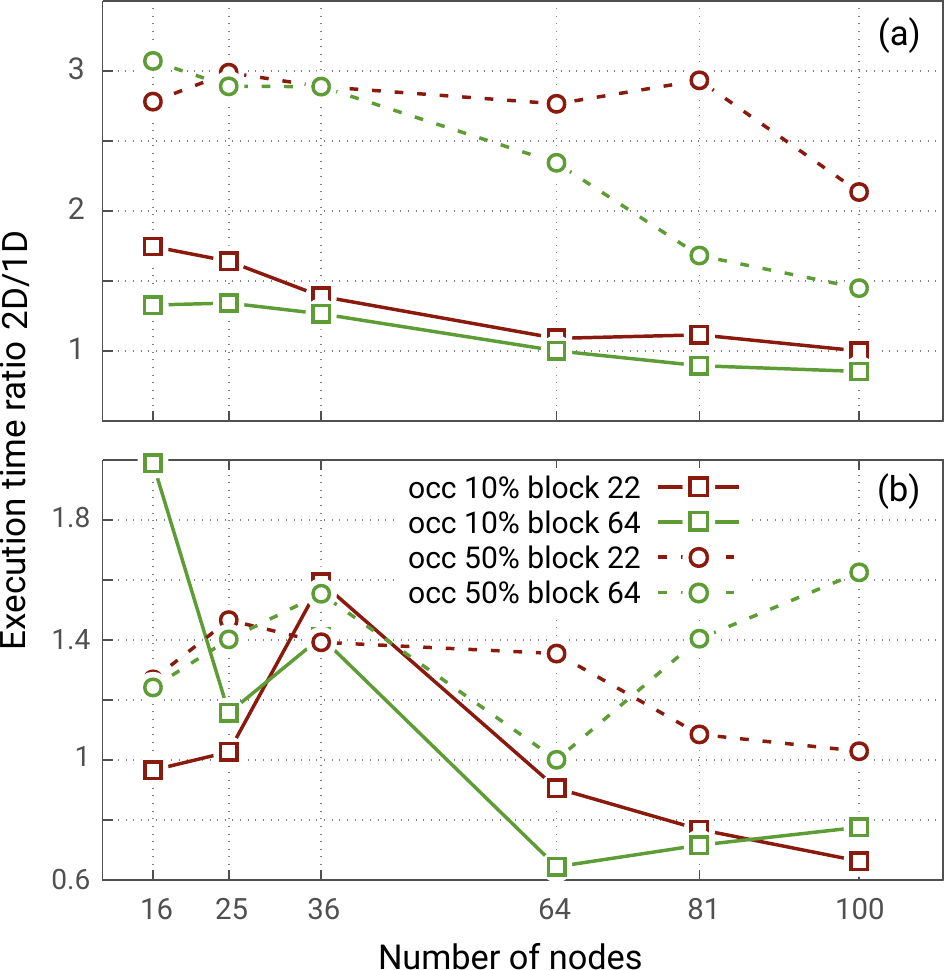}
\caption{Execution time ratios for two types of considered rectangular matrix multiplications on a 1D grid in comparison with the regular 2D Cannon algorithm. (a) The first type, $S_C \ll \{S_A, S_B\}$, shows significant improvement for higher occupancy (50\%) and less pronounced for the lower one (10\%).
(b) The second type, $S_B \ll \{S_A, S_C\}$, shows moderate to high speedup for higher occupancy (50\%) and poor to moderate behavior for the lower occupancy (10\%). In both cases, the benefit of the implementation reduces with the number of nodes, as expected.}
\label{fig:ratio}
\end{figure}

The results of the tests are presented in Figure~\ref{fig:ratio}b.
Overall, the new implementation gives a speed-up of up to 20\% with respect to the Cannon algorithm for high occupancy ($>10\%$) or up to 100\% for the matrices close to dense ($\sim 50\%$). The time for the redistribution and the overhead introduced by the virtual grid creation limits the benefit of the new implementation. For the same reasons, the benefit of the new algorithm is negligible or even worse for low occupancy.

\section{Sparse Tensor Algebra Implementation}
\label{sec:tensor}

\DBCSR was originally developed to enable linear scaling electronic structure methods that are mainly based on the multiplication of sparse square matrices. Similar strategies employing sparse data can also be employed for methods beyond density functional theory that provide better accuracy at significantly higher computational costs than Kohn-Sham density functional theory. In the case of the electron correlation methods MP2 and RPA, the canonical implementation scales at least quartic with system sizes, thereby preventing the study of large systems (hundreds to thousands of atoms). An initial \DBCSR-based cubic scaling implementation of RPA was reported in Ref.~\cite{cubic-rpa}, enabling calculations of thousands of atoms. Here we report strategies to optimize and generalize this initial implementation by extending the \DBCSR library to multi-dimensional tensors. A generalized implementation of tensor operations in \DBCSR instead of specialized implementations in the application code is desirable to manage code complexity and to easily extend the current implementation to other methods such as Hartree-Fock exchange or GW. The formalism of our RPA implementation was already described in Ref.~\cite{cubic-rpa} and here we emphasize the general characteristics of the tensor operations appearing in this and similar methods. We describe the requirements we pose for a tensor framework that should provide all relevant tensor operations in a general API.

As in \DBCSR, the sparsity of the tensors is based on the representation of molecular orbitals in terms of a localized atom-centered basis. A blocked sparsity pattern is equally important for the tensor implementation to efficiently incorporate sparse data while keeping the significant overhead for the handling of sparse indices low. Tensors can have arbitrary ranks, most relevant are tensors with ranks between 2 and 4. The main operation is tensor contraction where a sum over one or more indices of two tensors is performed. Our implementation is based on the property that tensor contractions are isomorphic to matrix-matrix multiplications \cite{solomonik2015communication}. This allows us to implement tensor contraction by mapping tensors to matrices -- the contraction is then internally performed by a call to the existing implementation of sparse matrix-matrix multiplication.

Recasting tensor contraction in terms of matrix-matrix multiplication imposes some requirements on the distribution and the matrix representation of the tensors, most importantly that one matrix dimension represents the indices to sum over and the other matrix dimension represents all other indices. If these requirements are not met, conversion steps are required before and after matrix-matrix multiplication which involves the redistribution of all tensor data. In order to avoid these relatively expensive redistribution steps, the tensor API gives the caller tight control over the distribution and the matrix representation of tensors, such that tensors can be created in a compatible layout and the redistribution step can be skipped in a tensor contraction. Data redistribution is then only strictly needed if a tensor appears in multiple contraction expressions involving sums over different indices.

While this approach of mapping tensors to matrices allows for an implementation of tensor operations as thin layers around an existing matrix library, the resulting matrices are problematic since one dimension is much larger than the other dimension. For the example of a 3 rank tensor with size $N\times N \times N$, two tensor dimensions are mapped to one matrix dimension such that one matrix dimension grows quadratically with the size of the other dimension. The \DBCSR library must thus be extended in a way that it can efficiently store and multiply tall-and-skinny matrices contrary to the previous target of square matrices.

One limitation of the \DBCSR matrix format is the index data replicated on all MPI ranks which contain information about block sizes and the distribution of blocks along each of the matrix dimensions. If the size of the matrix index corresponds to the number of atoms $N$ in a system, this limits the scalability of \DBCSR to a few tens of millions of particles \cite{joost1M}. For 3-rank tensors where the largest matrix dimension grows as $N^2$, this limit is already hit at a few thousand atoms, representing a much bigger issue in practice. Thus an extension to the \DBCSR matrix format must be provided to store large tensors without exhausting memory due to replicated index data. Another challenge is to multiply tall-and-skinny matrices communication-efficiently, where the algorithm described in section \ref{sec:rectangular} comes into play.

Our requirements for memory-efficient storage and communication-efficient multiplication can both be met by dividing the largest matrix dimension, resulting in smaller and approximately square submatrices that can be handled by \DBCSR. The storing of the full matrix index and the multiplication acting on submatrices are managed by an in-between tall-and-skinny matrix layer on top of \DBCSR that serves as a basis for the tensor implementation. The tall-and-skinny matrix layer is designed in a way that the index data is not explicitly stored but provided by externally defined function objects, to avoid the above-mentioned limitation of the \DBCSR format. The matrix index is thus handled in the tensor layer and is calculated on the fly from the tensor index. Due to the fully distributed sparse data layout, the matrix index calculation happens only when accessing a locally present non-zero block and does not add any overhead.

The main difference between the implementation of tall-and-skinny matrix multiplication and the one implemented in \DBCSR internally (see section \ref{sec:rectangular}) is that instead of relying on a linear process grid, the grid may have arbitrary dimensions. The submatrices are obtained on MPI subgroups by dividing any of the two grid dimensions by an arbitrary factor. This ensures that an optimal split factor can always be chosen, independently of the total number of processes, for any grid dimensions. Thus $n$-rank tensors can be represented on an arbitrary $n$-dimensional process grids where the grid dimension should be chosen as balanced as possible for a load-balanced distribution of data. The multiplication algorithm for contraction can then be run directly without additional costly redistribution steps (for tall-and-skinny matrices, the bandwidth cost of redistributing data exceeds the bandwidth cost for the multiplication \cite{carma}).

\section{Conclusion}
\label{sec:conclusion}

We have presented a new implementation for the rectangular matrix-matrix multiplication algorithm in the \DBCSR library that is able to speed-up the execution up to 3x for matrix sizes and occupancy values of $10\%-50\%$ which are often present in \CPTWOK calculations. We have described the newly developed tensor operations that generalize the \DBCSR library to multidimensional tensor contraction for low-scaling electronic structure methods beyond density functional theory. These functionalities are the basic building block for the \CPTWOK quantum chemistry and solid-state physics software package.

\section*{Acknowledgments}
We thank Shoshana Jakobovits (CSCS Swiss National Supercomputing Centre) for her work on the \LIBCUSMM code. 
This work was supported by grants from the Swiss National Supercomputing Centre (CSCS) under projects S238 and UZHP and received funding from the Swiss University Conference through the Platform for Advanced Scientific Computing (PASC).

\bibliographystyle{unsrt}
\bibliography{bib}

\end{document}